\documentclass[10pt,letterpaper,twocolumn]{article} 

\usepackage{ol2}
\usepackage[draft]{hyperref}
\usepackage{graphicx}
\usepackage{amssymb}
\usepackage{amsmath}
\usepackage{epstopdf}
\usepackage{ulem}
\usepackage[squaren,Gray,cdot]{SIunits}

\begin{document}

\twocolumn[

\title{A reversible optical memory for twisted photons}

\author{L. Veissier$^{1}$, A. Nicolas$^{1}$, L. Giner$^{1}$, D. Maxein$^{1}$, A.S. Sheremet$^{2}$, E. Giacobino$^{1}$, J. Laurat$^{1,*}$}

\address{
$^1$Laboratoire Kastler Brossel, Universit\'{e} Pierre et Marie Curie, Ecole Normale Sup\'{e}rieure, CNRS, \\4 place Jussieu, 75252 Paris Cedex 05, France \\
$^2$Department of Theoretical Physics, State Polytechnic University, 195251, St.-Petersburg, Russia
\\
$^*$Corresponding author: julien.laurat@upmc.fr
}

\begin{abstract}
We report on an experiment in which orbital angular momentum of light is mapped at the single-photon level into and out of a cold atomic ensemble. Based on the dynamic electromagnetically-induced transparency protocol, the demonstrated optical memory enables the reversible mapping of Laguerre-gaussian modes with preserved handedness of the helical phase structure. The demonstrated capability opens the possibility to the storage of qubits encoded as superpositions of orbital angular momentum states and to multi-dimensional light-matter interfacing. 
\
\end{abstract}
]

\noindent The orbital angular momentum (OAM) of light has raised a great deal of applications \cite{Allen,Torres}, ranging from the trapping of particles, the driving of optical micromachines to applications in astrophysics studies. In quantum optics, twisted single-photons \cite{Leach02} have also been identified as promising carriers for the implementation of various quantum information protocols \cite{ZeilingerNature01,PadgettQInfo04,ZeilingerQInfo06}. They indeed offer the possibility to encode and process information in high-dimensional Hilbert spaces \cite{Molina, Dada, Fickler} and could enable to reach higher efficiencies and enhanced information capacity.

For these interesting features to be extended to quantum networks involving light-matter interfaces \cite{Kimble08}, various experiments based on the coherent interaction of light beams carrying OAM with atomic systems have been performed. For instance, entanglement of OAM states between a collective atomic excitation and a photon has been demonstrated \cite{Kozuma06} based on an adapted version of the measurement-induced Duan-Lukin-Cirac-Zoller protocol. A critical capability for further applications is also the ability to reversibly map OAM into and out of atomic memories. In the past few years, storage and retrieval of light beams carrying OAM have indeed been demonstrated \cite{Davidson07,Tabosa09}, but these seminal works involved bright beams. Here, we report the realization of a multimode optical memory enabling the conservation of the orbital angular momentum, for the first time at the single-photon level.

In this work, we specifically focus on the Laguerre-gaussian (LG) modes LG$_{p}^l$ with a radial index $p=0$ and a winding number $l=\pm 1$. Such modes exhibit a doughnut-shape intensity distribution and have a phase with an azimuthal dependence $e^{il\theta}$ quantified by its circulation $l$ around the beam axis. They
carry an orbital angular momentum $l$ per photon in units of $\hbar$, defining the handedness of the helical wavefront. 

The experimental realization proceeds in the following way. First, LG modes are generated using a spatial light modulator. These modes, which are weak coherent states at the single-photon level, are then mapped into an ensemble of cold atoms using the dynamic electromagnetically-induced transparency (EIT) protocol
\cite{LukinRMP03,FleischhauerRMP05,Lvovsky}. This method relies on
three-level atoms in a $\Lambda$ configuration interacting with a strong control field on one of
the optical transitions. This interaction leads to transparency for a resonant field on the other transition,
which would otherwise be absorbed. In addition to transparency, the
frequency dependence of the refractive index associated with EIT
causes a dramatic group velocity reduction, leading to slow light \cite{HauNat99}. By switching off adiabatically the control field, the light pulse can be converted into a
coherent superposition of the two atomic ground states. The signal
quantum properties, including its OAM, are expected to be transferred
to the atomic medium
and can be stored for a programmable time, limited by the lifetime
of the atomic coherence. Turning on the control field again enables
to retrieve the signal pulse. Finally, after this reversible storage, the retrieved light is characterized via mode discriminators and avalanche photodiodes, as explained thereafter.

\begin{figure*}[t!]
\centerline{\includegraphics[width=1.8\columnwidth]{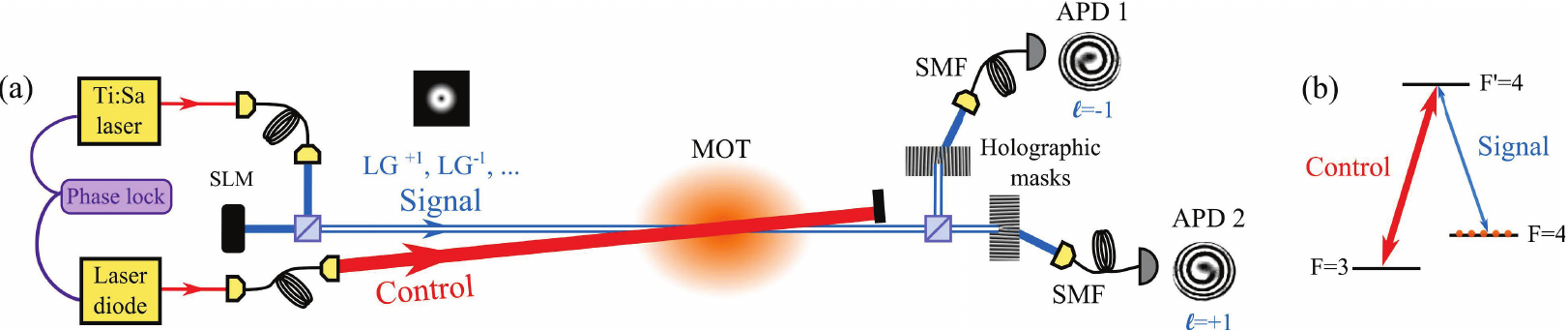}}
\caption{(a) Experimental setup. A signal pulse consisting in a weak coherent state at the single-photon level is orbitally shaped by
reflection on a spatial light modulator (SLM) and then mapped into an ensemble of cold cesium atoms. After a user-defined delay, the collective atomic excitation is readout. The retrieved light is analysed via splitting on a 50/50 beam-splitter and an OAM mode projection on each path. This projection results from an OAM addition (for the $l=-1$ path) or subtraction (for the $l=+1$ path) through blazed fork holograms and a subsequent coupling into single-mode fibers. The coupled light is detected by avalanche photodiodes (APD). (b) Relevant energy diagram.} \label{Exp_setup}
\end{figure*}

Indeed, unlike for bright beams, the detection of twisted single-photons cannot be performed by imaging on a CCD camera. Several techniques have thus been developed for single-photon OAM detection\cite{Padgett2SLM10,Leach02,ZeilingerNature01,ZeilingerJOptB02,PadgettOAMModeSorter12}. Here we have chosen to work with a combination of a computer-generated hologram and a single-mode fiber to implement a mode discriminator \cite{ZeilingerNature01,ZeilingerJOptB02}. The principle is the following. The light to characterize is first diffracted with great efficiency ($>80\%$) into the first order of a blazed fork phase hologram, as described in \cite{ZeilingerJOptB02}. Given the orientation of the centered dislocation, one unit of OAM is added or subtracted. The resulting mode is then directed into a standard single-mode optical fiber. As only the fundamental TEM$_{00} $ gaussian mode is efficiently coupled into such a fiber, this scheme allows to extract only one OAM component of the light beam. All other components are filtered out. In our experiment, the output mode is actually split on a $50/50$ beam splitter and one discriminator is installed in each of the two output paths so that we can quantify both the LG$^{+1}$ and LG$^{-1}$ components of the retrieved light. We achieve a high distinction ratio of 17 dB and 23 dB respectively. 

The experimental setup is sketched in Fig. \ref{Exp_setup}. The atomic medium consists in
cesium atoms in a magneto-optical trap (MOT). The three-level
$\Lambda$-type system in which EIT takes place involves the two ground
states, $|g\rangle=|6S_{1/2},F=4\rangle$ and
$|s\rangle=|6S_{1/2},F=3\rangle$, and one excited state
$|e\rangle=|6P_{3/2},F=4\rangle$. All the
atoms are initially prepared in $|g\rangle$. The memory is operated in sequence, on a free expanding cloud. The
magnetic field and the trapping beams are turned off after a 19 ms
cloud buildup period. The initial optical depth of the cold atomic ensemble is set around 15. In order to increase the coherence time, the magnetic field has been canceled down to 5 mG via microwave spectroscopy.

The control field is resonant
with the $|s\rangle$ to $|e\rangle$ transition while the signal field
is resonant with the $|g\rangle$ to $|e\rangle$ transition.  Signal and control fields are
generated respectively from a Ti:Sapphire laser and from an extended
cavity laser diode, and they are
phase-locked at the atomic hyperfine frequency. The control field is
horizontally-polarized, with a 200 $\mu$m waist in the
MOT, a power of 15 $\mu$W and a \unit{2}{\degree} angle
relative to the direction of the signal beam. The signal field is
vertically-polarized, with  a waist of 50 $\mu$m.

\begin{figure*}[h!]
\centerline{\includegraphics[width=1.3\columnwidth]{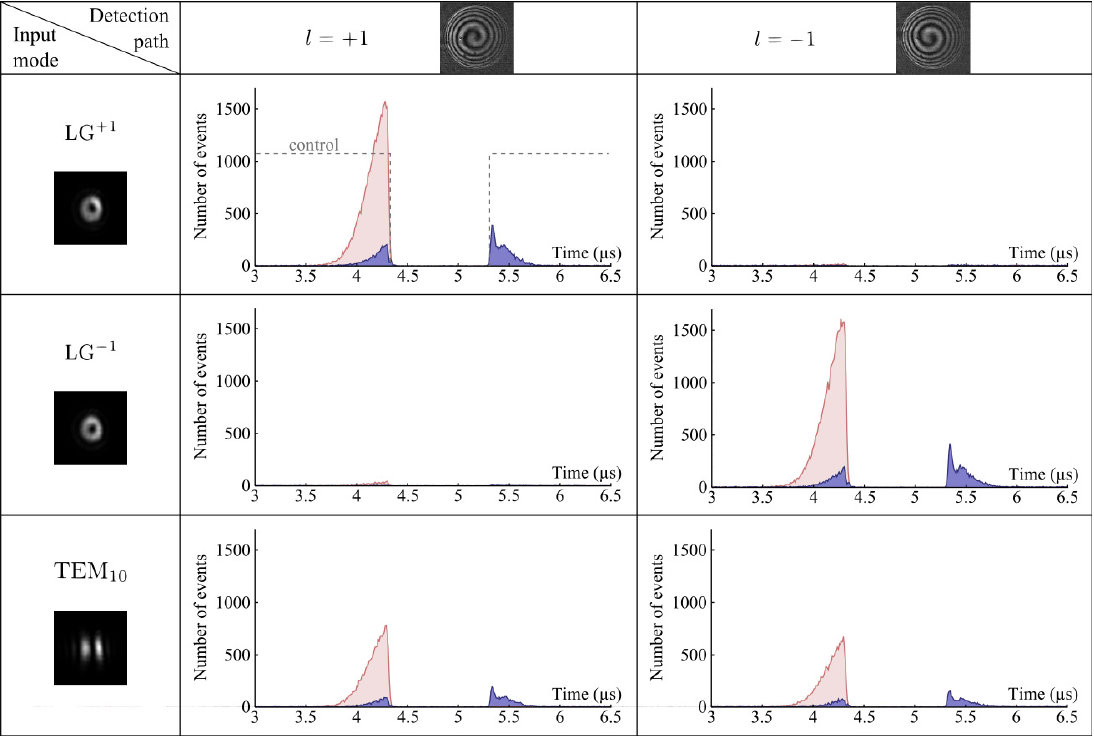}}
\caption{Storage and retrieval of LG$^{+1}$, LG$^{-1}$, and TEM$_{10}$ modes at the single-photon level. For the two simultaneous detection paths, the
number of counts is represented as a function of time. Light red lines represent the incoming signal pulses recorded without atoms and dark blue
lines correspond to memory measurements. The events around the incoming pulses show the leakage of the signal while the later events correspond to the readout. The dotted line in the first plot shows the timing of the control pulse. Each curve results from $5.10^5$ repetitions of the experiment.} \label{TableOfResults}
\end{figure*}

The signal pulses are temporally shaped into a half gaussian profile with a typical width of 0.5 $\mu$s and a mean number of photons per pulse of 0.6. Photons are detected by a pair of avalanche photodiodes (APD, model SPCM-AQR-14-FC) placed after the discriminators on each path. Events from both APDs are simultaneously recorded
with a time resolution of 10 ns. In addition to the intrinsic spatial filtering due to the signal/control angle, the control beam is additionally filtered out in polarization. A 100 dB attenuation is therefore obtained. The memory sequence is repeated 100 times every 25 ms. 

We next turn to the experimental results. Figure ~\ref{TableOfResults} shows the storage and retrieval of different OAM states. Both detection paths ($l=+1$ and $l=-1$) are displayed. The first two lines give the results respectively for an impinging LG$^{+1}$  mode and for a LG$^{-1}$ mode. As the pulse propagates into the medium, the control field is turned off in 100 ns and the photonic information is mapped into collective atomic excitations. After a chosen delay of 1 $\mu$s, the control field is switched on again and the atomic excitations are converted back to a propagating light pulse. Blue lines in Fig. \ref{TableOfResults} give the recorded single-photon detection events: counts around the incoming pulse correspond to leakage and the later events represent the retrieved pulse. As can be seen, detection events in the detection path with $l$ opposite to the one of the incoming mode are negligible, within the distinction ratio. For both modes, the overall storage and retrieval efficiency is ($16 \pm 2$$\%$) (given by the ratio of events detected for the retrieved pulse and the events for the incoming pulse without atoms). These results confirm the preservation of orbital angular momentum at the single-photon level upon the reversible mapping process demonstrated here.

As shown in the last line of Fig. \ref{TableOfResults}, we also stored and retrieved a TEM$_{10}$ mode, i.e. an equal-weight superposition of LG$^{+1}$ and LG$^{-1}$ modes. In the ideal case, equal number of events are expected in the two detection paths. Experimentally, a 9$\%$ imbalance is observed between the two paths whatever the observed pulses, i.e. for the reference and the retrieved pulses. This imbalance can by explained by the fact that the mode discriminators have different behaviors with respect to the higher order radial modes LG$_{p>0}$. Indeed, due to our mode preparation with a phase only modulation, these unwanted modes contribute more in our experimentally generated HG beams than in the LG beams. Within our experimental precision, the achieved storage and retrieval efficiency  is the same here for each LG components than for the individual storage presented previously.  Further works should include the reconstruction of a full density matrix, including coherences not measured here, and the calculation of the process fidelity. 

In conclusion, we have reported the storage and retrieval of light pulses carrying orbital angular momentum in an EIT setting. Importantly, these experiments have been performed at the single-photon level, opening the way to the demonstration of a quantum memory for qubit encoded as OAM superpositions. Extensions to higher alphabets for the realization of high-dimensional quantum information networks is also work for future.\\

% Ack
%===
We thank A. Zeilinger, R. Fickler, M.J. Padgett and D. Tasca for their assistance with fork holograms and SLM. This work is supported
by the ERA-Net CHIST-ERA (QScale), by the ERA-Net.RUS (Nanoquint), by the ICT/FET HIDEAS project, by the CNRS-RFBR collaboration
(CNRS 6054 and RFBR P2-02-91056) and by the ERC starting grant HybridNet. A. Sheremet acknowledges the support from ``Dynasty" and A. Nicolas from DGA. J. Laurat is a member of the Institut Universitaire de France.

\end{document}